\documentclass[pra,aps,reprint,a4paper,superscriptaddress,floatfix,longbibliography]{revtex4-2}
\usepackage{times}
\usepackage{graphicx}
\usepackage{epsfig}
\usepackage{amsfonts}
\usepackage{amsmath}
\usepackage{amssymb}
\usepackage{dsfont}
\usepackage{amsthm}
\usepackage{color}
\usepackage{comment}
\usepackage{braket}
\usepackage{physics}
\usepackage{multirow}
\usepackage{diagbox}
\usepackage{multirow}
\usepackage{makecell}
\usepackage{enumerate}

\usepackage{float}
\usepackage{rotating}
\usepackage[short,c3]{optidef}
\usepackage[linesnumbered,ruled]{algorithm2e}
\usepackage{scalerel}[2016/12/29]

\usepackage[colorlinks=true,linkcolor=blue,citecolor=blue,urlcolor=blue]{hyperref}

\begin{document}

\renewcommand{\arraystretch}{1.3}


\title{Reply to Pavi\v{c}i{\'c}'s `Comment on ``Optimal conversion of Kochen-Specker sets into bipartite perfect quantum strategies''\,' (arXiv:2502.13787)}


\author{Stefan Trandafir}
\affiliation{Departamento de F\'{\i}sica Aplicada II, Universidad de Sevilla, E-41012 Sevilla, Spain}

\author{Ad\'an~Cabello}
\email{adan@us.es}
\affiliation{Departamento de F\'{\i}sica Aplicada II, Universidad de Sevilla, E-41012 Sevilla,
Spain}
\affiliation{Instituto Carlos~I de F\'{\i}sica Te\'orica y Computacional, Universidad de
Sevilla, E-41012 Sevilla, Spain}


\begin{abstract}
According to Pavi\v{c}i{\'c}, Kochen and Specker's 117-observable set is not a ``Kochen-Specker set''.
By the same reason, in arXiv:2502.13787, Pavi\v{c}i{\'c} claims that 10 statements in our paper ``Optimal conversion of Kochen-Specker sets into bipartite perfect quantum strategies'' [Phys. Rev. A 111, 022408 (2025)] are ``wrong''.
In all cases, Pavi\v{c}i{\'c}'s claims are based on the fact that he is assuming a different definition of Kochen-Specker (KS) set. Adopting a terminology used by, e.g., Larsson, the sets that Pavi\v{c}i{\'c} call KS sets can be called ``extended'' KS sets, since they are constructed by adding observables to the ``original'' KS sets. For example, Pavi\v{c}i{\'c} adds 75 observables to the original 117-observable KS set. Beyond terminology, there are fundamental physical reasons for focusing on the original KS sets. One reason is that, for experimentally observing quantum state-independent contextuality, there is no need to measure the observables added in the extended sets. Another reason is that, to produce bipartite perfect quantum strategies, or correlations in a face of the nonsignaling
polytope with no local points, or correlations with nonlocal content 1, the two parties do not need to measure any of the observables added in the extended sets. We also respond to other claims made by Pavi\v{c}i{\'c} about our work.
\end{abstract}

\maketitle


According to Pavi\v{c}i{\'c} \cite{Pavicic:2025}, our paper \cite{TrandafirCabello:2025} contains ``a number of errors, inconsistencies, and inefficiencies''. In the following, we respond to these claims.

Regarding ``errors'', in Secs.\ B and C of \cite{Pavicic:2025}, Pavi\v{c}i{\'c} claims that 10 statements in \cite{TrandafirCabello:2025} are ``wrong''. In all cases, the claims are based on the fact that he is assuming a different definition of KS set. Specifically, in \cite{TrandafirCabello:2025}, we are using the ``original'' definition (as it is called in, e.g., \cite{Larsson:2002EPL}), namely, the one used by Kochen and Specker in \cite{Kochen:1967JMM}.

{\em Definition.---}A KS set in finite dimension
$d \ge 3$ is a finite set of rank-one projectors (or rays or directions) $V$ in a Hilbert space
of dimension $d$ which does not admit an assignment $f : V \rightarrow
\{0, 1\}$ satisfying: (I) $f(u) + f(v) \le 1$ for each pair of orthogonal projectors
$u, v \in V$. (II) $\sum_{u \in b} (u) = 1$ for every set $b \subset V$ of mutually orthogonal projectors whose sum is the identity.

Larsson~\cite{Larsson:2002EPL} introduced an alternative definition in which condition (I) is removed. Larsson refers to the resulting sets as ``extended'' KS sets \cite{Larsson:2002EPL}, since they are constructed by adding observables to the original KS sets. Every extended KS set is an original KS set, but not the other way around. Neither Kochen and Specker's 117-observable set \cite{Kochen:1967JMM}, nor Sh\"utte's 33-observable set \cite{Bub:1996FP}, nor Peres' 33-observable set \cite{Peres:1991JPA}, nor Penrose's 33-observable set \cite{Penrose:2000}, nor Conway and Kochen's 31-observable set \cite{Peres:1993} are extended KS sets. However, all of them are original KS sets. The same holds true for the KS sets in \cite{TrandafirCabello:2025} that Pavi\v{c}i{\'c} says are wrong. There is nothing wrong, we simply adopted the above definition (which is clearly stated in \cite{TrandafirCabello:2025}!), while Pavi\v{c}i{\'c} is assuming a different definition.

Beyond terminology, there are fundamental physical reasons for focusing on the original KS sets. One reason is that, for experimentally observing quantum state-independent contextuality, there is no need to measure the observables added in the extended KS sets; the observables in the original KS sets suffice. The reason is the following.

{\em Theorem 1.---} Given any original KS set ${\cal K}$ in dimension $d$, there is a noncontextuality inequality such that any quantum state in dimension $d$ violates it using only the elements of ${\cal K}$. 

To prove it, we use the following Lemma taken from \cite{xu2023stateindependent}.

{\em Lemma.---}Given a finite set of observables $\{\Pi_i\}$, with possible results $0$ or $1$, and graph of compatibility $G$ (in which each $\Pi_i$ is represented by a vertex $i \in V$ and there is an edge $(i,j) \in E$ if $\Pi_i$ and $\Pi_j$ are compatible), the following inequality holds for any noncontextual hidden-variable (NCHV) model:
\begin{equation}\label{eq:gnci}
{\cal W} := \sum_{i \in V} w_i\ P_i - \sum_{(i,j) \in E} w_{ij} P_{ij} \overset{\rm NCHV}{\le} \alpha(G,\vec{w}),
\end{equation}
where $\vec{w}=\{w_i\}_{i \in V}$ is a set of positive weights for the vertices of $G$,
$w_{ij} \ge \max{(w_i,w_j)}$, $P_i = P(\Pi_i=1)$ is the probability of obtaining outcome $1$ when measuring observable $\Pi_i$, $P_{ij}=P(\Pi_i=1,\Pi_j=1)$ is the probability of obtaining outcomes $1$ and $1$ when measuring $\Pi_i$ and $\Pi_j$, and $\alpha(G,\vec{w})$ is the weighted independence number of $G$ with vertex weight vector $\vec{w}$.

That at least one of the above inequalities is violated by any original KS set can be proven as follows. 
Given a $d$-dimensional original KS set ${\cal K}$, let $G$ be its graph of compatibility. Let $N$ be the number of cliques of size $d$ in $G$. Notice that one may disregard from ${\cal K}$ all projectors corresponding to vertices that are not in a clique of size $d$ and obtain an original KS set where all projectors correspond to vertices that are in a clique of size $d$.
Now consider the functional ${\cal W}$ given by Eq.~\eqref{eq:gnci} such that $w_i$ is the number of cliques of size $d$ that cover vertex $i$. Then, for any original KS set, 
\begin{equation}
\alpha(G,\vec{w}) < N.
\end{equation}
However, for any quantum state in dimension $d$, 
\begin{equation}
{\cal W}=N.
\end{equation}
For example, a noncontextuality inequality violated by any qutrit state using only Peres' 33-observable set \cite{Peres:1991JPA} is presented in \cite{xu2023stateindependent}[Supplemental Material, Table VI].

More importantly, to produce bipartite perfect quantum strategies ---which is the goal of \cite{TrandafirCabello:2025}--- the distant parties do not need to measure any of the observables added in an extended KS set; the observables in the original KS set suffice. The reason is the following.

{\em Theorem 2.---}Given any original KS set ${\cal K}$ in dimension $d$, there is a bipartite perfect quantum strategy with quantum advantage in which the two parties only measure elements of ${\cal K}$. 

The algorithm in \cite{TrandafirCabello:2025} produces such a bipartite perfect quantum strategy.

Taking into account the results in \cite{Liu:2024}, Theorem 2 can be reformulated as follows:

{\em Theorem 2b.---}Given any original KS set ${\cal K}$ in dimension $d$, there is a bipartite correlation in a face of the nonsignaling polytope with no local points produced by only measuring the elements of ${\cal K}$. 

{\em Theorem 2c.---}Given any original KS set ${\cal K}$ in dimension $d$, there is a bipartite correlation with nonlocal content 1 produced by only measuring the elements of ${\cal K}$. 

As pointed out in \cite{Cabello:2025}: ``while KS sets were initially important because of state-independent contextuality (\ldots), now we know that they are not needed for state-independent contextuality (\ldots). However,
(\ldots) KS sets are necessary (\ldots) for bipartite perfect quantum strategies. That is, KS sets are fundamental in {\em Bell nonlocality} (\ldots). More generally, (\ldots) KS sets are crucial for groundbreaking results in
quantum computation, information, and foundations.'' However, by KS sets, we mean the original KS sets. Of course, any set of rank-one projectors that contains an original KS set is also an original KS set.

Pavi\v{c}i{\'c} also states that our representation of the KS sets in \cite{TrandafirCabello:2025} is ``inefficient''. Our aim in \cite{TrandafirCabello:2025} was to provide information in a clear readable manner. In most cases, we explicitly gave the vectors and orthogonal bases (as it is a standard practice \cite{Peres:1991JPA,Peres:1993,Bub:1996FP}). In one instance, for the purposes of simplifying the presentation, we gave the orthogonalities explicitly and referred to the vectors implicitly. In all cases, we provide sufficient information so that anyone may reproduce our results.

In addition, in Secs.\ E and F of \cite{Pavicic:2025}, Pavi\v{c}i{\'c} provides references for KS sets in dimensions 5 and 7 having the same number of rank-one projectors as the KS sets we use in \cite{TrandafirCabello:2025}, but, in the case of dimension $7$, with fewer orthogonal bases. However, if one uses any of the sets that are explicitly given in Pavi\v{c}i{\'c}'s articles as starting point of our algorithm, one does not obtain bipartite perfect strategies with smaller input cardinality. In this sense, these sets are not more efficient than the ones used in \cite{TrandafirCabello:2025}. Of course, it may be the case that, starting with different KS sets, the algorithm produces a bipartite perfect quantum strategy with a smaller input cardinality. However, so far, all the records in dimensions from 3 to 8 are in \cite{TrandafirCabello:2025}.

Summing up, we believe that Pavi\v{c}i{\'c}'s claims about our work having errors are unfounded. We take this opportunity to encourage researchers to try to further reduce the input cardinalities and identify the smallest values allowed by quantum theory for every dimension.


%


\end{document}